\documentclass[pra,aps,twocolumn,showpacs]{revtex4}
\usepackage{epsfig}
\usepackage{amsmath}

\newcommand{\eh}{\mathrm{e}}
\newcommand{\ii}{\mathrm{i}}

\newcommand{\br}{\mathbf{r}}

\newcommand{\diff}{\mathrm{d}}

\begin{document}

\title{Dynamics of vortices in weakly interacting Bose-Einstein condensates}

\author{Alexander Klein and Dieter Jaksch}

\affiliation{Clarendon Laboratory, University of Oxford, Parks Road,
Oxford OX1 3PU, United Kingdom, \\
and Keble College, Parks Road, Oxford OX1 3PG, United Kingdom}

\author{Yanzhi Zhang and Weizhu Bao}

\affiliation{Department of Mathematics and Center for Computational
Science {\rm \&} Engineering, National University of Singapore,
Singapore 117543 }

\date{\today}

\begin{abstract}

We study the dynamics of vortices in ideal and weakly interacting
Bose-Einstein condensates using a Ritz minimization method to solve
the two-dimensional Gross-Pitaevskii equation. For different initial
vortex configurations we calculate the trajectories of the vortices.
We find conditions under which a vortex-antivortex pair annihilates
and is created again. For the case of three vortices we show that at
certain times two additional vortices may be created, which move
through the condensate and annihilate each other again. For a
noninteracting condensate this process is periodic, whereas for
small interactions the essential features persist, but the
periodicity is lost. The results are compared to exact numerical
solutions of the Gross-Pitaevskii equation confirming our analytical
findings.

\end{abstract}



\maketitle


\section{Introduction}

Quantized vortices play an important role in verifying the
superfluid properties of quantum liquids such as Bose-Einstein
condensates (BECs) or degenerate Fermi gases. In weakly interacting
alkali gases condensate states containing a single vortex line were
first created using Raman transition phase-imprinting methods
\cite{Matthews-PRL-1999}. By rotating the system with a laser spoon
\cite{Madison-PRL-2000,Madison-JMO-2000}, vortex lattices containing
more than 100 vortices have been created
\cite{Abo-Shaeer-Science-2001,Raman-PRL-2001}, and by using
topological phase engineering methods \cite{Leanhardt-PRL-2002} it
is even possible to create multiply charged vortices. It is expected
that more complicated vortex clusters can be created in the future,
e.g., with the further development of phase-imprinting methods. Such
states would enable various opportunities, ranging from
investigating the properties of random polynomials
\cite{Castin-PRL-2006} to using vortices in quantum memories
\cite{Kapale-PRL-2005}. All of these developments stir a great
interest in the study of states with several vortices.

Recently, there were a number of investigations on the properties of
vortices in BECs. For three-dimensional condensates, several studies
on the dynamics of vortex lines have been done
\cite{Jackson-PRA-1999,Feder-PRL-2001}. The generation and dynamics
of vortices in a toroidal condensate have been investigated in
Ref.~\cite{Martikainen-PRA-2001}, whereas detailed numerical studies
of the optical generation of vortices in pancake-shaped condensates
have been carried out in \cite{Andrelczyk-PRA-2001}. The
manipulation of vortices such as charge conversion by external
potentials has been discussed in Ref.~\cite{Perez-Garcia-PRA-2007}.
Further numerical studies revealed that for condensates with a
strong nonlinearity there exist several configurations of vortices
which are stable
\cite{Crasovan-PRA-2003,Moettoenen-PRA-2005,Pietilae-PRA-2006}.
Analytical expressions for the angular momentum and the energy of a
vortex-antivortex configuration in a BEC have been obtained
\cite{Zhou-PRA-2004} using the Thomas-Fermi approximation. For
strongly nonlinear condensates analytical solutions were derived by
splitting the wave function into a region close to the vortex core
and one far away from the vortex, where the hydrodynamic properties
of the condensate are an essential feature
\cite{Rubinstein-PysicaD-1994,Pismen-1999,Anglin-PRA-2002}. In
contrast, here we concentrate on the dynamical properties of vortex
configurations in weakly interacting condensates, and solutions for
the whole spatial regime are obtained. In a noninteracting
condensate the vortices behave similar to those created in an
optical beam using holograms
\cite{Terriza-OptLett-2001,Roux-OC-2004}, where the time evolution
in the BEC corresponds to the spatial evolution of the laser beam.
As we will show, the interaction between the BEC atoms changes the
behavior of the vortex dynamics considerably.

In this paper, we make use of the Gross-Pitaevskii equation (GPE),
also known as the nonlinear Schr{\"o}dinger equation, which is known
to be a valid description of the mean-field dynamics of a BEC at
zero temperature. We consider a harmonic trap with tight confinement
along one direction, such that the condensate is effectively two
dimensional. For the case of an ideal, i.e., noninteracting BEC, the
dynamics of the vortices is solved analytically yielding the
essential features of the time evolution. For a small interaction
within the condensate, we make use of the Ritz method in order to
get analytical estimates of the dynamics. These estimates are
compared to exact numerical solutions of the GPE using the
time-splitting spectral method (TSSP), which is explicit,
unconditionally stable and spectrally accurate in space. Details of
the numerical method are described in
Refs.~\cite{Bao-JCP-2003,Bao-SIAM-2003,Bao-MMMAS-2005}.

This paper is organized as follows. In Sec.~\ref{Sec:Model} we
introduce the model under investigation and define the general
initial states of the vortex configurations. In Sec.~\ref{Sec:Ideal}
we discuss the dynamics of vortices in an ideal condensate as a
background for the results of Sec.~\ref{Sec:IA}, where a detailed
investigation of the dynamics of a single vortex, a vortex pair, a
vortex dipole, and a vortex tripole are presented. We conclude in
Sec.~\ref{Sec:Conclusion}. An Appendix contains some more details on
the evolution of a vortex tripole for the noninteracting BEC.


\section{The model \label{Sec:Model}}

In this work, we consider a Bose-Einstein condensate (BEC) in a
radially symmetric trap $V_t(x,y,z)=\frac{1}{2}m_b[\omega
(x^{2}+y^{2})+\omega _{z}z^{2}]$ with $\omega _{z}\gg \omega$ the
trap frequencies in axial and radial direction, respectively, and
$m_b$ the mass of the BEC atoms. We assume a tight confinement in
axial direction such that $\hbar \omega_z \gg k_{\mathrm{B}} T$,
where $k_{\mathrm{B}}$ is Boltzmann's constant and $T$ is the
temperature of the BEC, as well as $\hbar \omega_z \gg  g n_0$, with
$n_0$ the density of the BEC in the center of the trap and $g$ the
interaction strength within the BEC, given by $g = 4 \pi \hbar^2
a_s/m_b$, with $a_s$ the $s$-wave scattering length. For
temperatures well below the critical temperature of the BEC and the
Berezinskii-Kosterlitz-Thouless transition temperature
\cite{Berezinskii-JETP-1972,Kosterlitz-JPC-1973}, a regime which is
in reach of current experiments
\cite{Hadzibabic-Nature-2006,Krueger-2007}, phase fluctuations occur
on scales which are typically larger than the size of the condensate
and the dynamics of the BEC is well described by the dimensionless
2D Gross-Pitaevskii equation \cite{Bao-JCP-2003}
\begin{equation} \label{Eq:GPE}
\ii\frac{\partial \psi }{\partial t}=\left[ -\frac{1}{2}\left(
\frac{\partial ^{2}}{\partial x^{2}}+\frac{\partial ^{2}}{\partial
y^{2}}\right) +\frac{1}{2} (x^2 + y^2) +\beta |\psi |^{2}\right]
\psi \,.
\end{equation}
Here, $\psi =\psi (x,y,t)$ is the normalized wave function of the
condensate with $\int |\psi (x,y)|^{2}\diff x \diff y=1$, and $\beta
=2 N a_s \sqrt{2 \pi \omega_z/\omega}\,/a_0$ characterizes the
interatomic interaction, defined in terms of the total number of
particles $N$ in the condensate. The above dimensionless quantities
are obtained by scaling the length by the harmonic oscillator length
$a_0=\sqrt{\hbar/m_b\omega }$, the time by $\omega^{-1}$, and the
energy by $\hbar \omega$.

We study the dynamics of $n$ vortices with topological charge $q_j =
\pm 1$ ($j = 1,2,...,n$), which are initially placed at positions
$\br_j=(x_j,y_j)$. For this purpose, we first need to calculate the
ground state $\psi_{\mathrm{gs}}(\br)$ of the GPE and the state
$\psi_{q}(\br)$ with a single vortex in the center of the trap. From
this, we extract the function $p_{q}(\br) =\psi_{q}(\br) /
\psi_{\mathrm{gs}}(\br)$, which describes a vortex in the BEC
background. The initial state for the vortices is then approximately
given by
\begin{equation} \label{Eq:InitalData}
  \psi(\br,t=0) = \alpha  \psi_{\mathrm{gs}}(\br)
   \prod_{j = 1}^n p_{q_j}(\br - \br_j)
    \,,
\end{equation}
where $\alpha$ is chosen such that the initial state is normalized
to 1. Unless otherwise stated, this normalization constant will be
dropped in the following. The above approximation holds for vortices
which are not too close to the edge of the condensate. A physical
realization of such states can be achieved either by stirring the
condensate \cite{Madison-PRL-2000,Madison-JMO-2000} or, in a more
controlled way, by phase imprinting methods
\cite{Matthews-PRL-1999,Leanhardt-PRL-2002} or by using light with
orbital angular momentum \cite{Kapale-PRL-2005}.


\section{Vortices in a noninteracting BEC \label{Sec:Ideal}}

To get an insight into the dynamics of vortices and as a background
against the results found for the weakly interacting condensate we
first focus on the noninteracting case $\beta = 0$, where the GPE
simplifies to a two-dimensional harmonic oscillator. The initial
state $\psi(\br,t=0)$ is expanded in terms of the solutions
\begin{equation}
\psi_{n,m}(\br,t) = \frac{\eh^{- \ii E_{n,m} t  -(x^2 + y^2)/2
}}{\sqrt{2^{n+m} n! m! \pi}} \,  H_n(x) H_m(y)
\end{equation}
of the harmonic oscillator, with the energies $E_{n,m} = 1+n+m$,
$H_n$ are Hermite polynomials, and integer numbers $n,m \geq 0$.
This also gives the time evolution of the vortex state. The
trajectories of the vortices are calculated by finding the zeros of
the wave function and checking if at these points the condensate has
a nonzero winding number.

For a single vortex of topological charge $q=+1$ initially located
at $(x_1,0)$ the expansion gives the wave function
\begin{equation}
  \phi_{\mathrm{v}}(x,y,t) =
  \exp\left(- \frac{x^2 + y^2}{2} - 2 \ii t \right)\,
  \left[ x + \ii y - \eh^{\ii t}x_1 \right] \,.
\end{equation}
The position of the vortex evolves in time on an exact circular
trajectory described by $x_{\mathrm{v}}(t) = x_1 \cos(t)$ and
$y_{\mathrm{v}}(t) = x_1 \sin(t)$.

For a vortex pair, that is two vortices with identical topological
charge $q = +1$ initially located at $(x_j,y_j)$, where $j=1,2$, the
expansion into the solutions of the harmonic oscillator shows that
the two vortices move independently from each other on trajectories
$x_{{\mathrm{vp}},j}(t) = x_j \cos(t) - y_j \sin(t)$ and
$y_{{\mathrm{vp}},j}(t) = x_j \sin(t) + y_j \cos(t)$, which are
exactly the same as for a single vortex. Due to the conservation of
the total topological charge $Q = \sum_j q_j$ there are always at
least two vortices present in the condensate, since a doubly charged
vortex is unstable \cite{Pitaevskii-2003}. From the absence of
additional vortices we conclude that the total energy of the system
is not large enough to allow the spontaneous creation of a
vortex-antivortex pair.

\begin{figure}
\includegraphics[width=8cm]{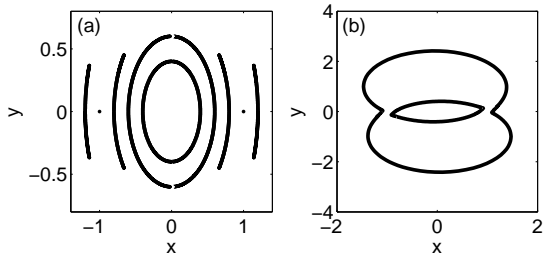}
\caption{(a) Trajectories for a vortex dipole for different initial
positions $(\pm x_0,0)$. From inside to outside $x_0 =
0.4,0.6,0.8,1,1.2$. For $x_0 = 0.4,0.6$, the two vortices annihilate
each other at certain times and reappear again, indicated by the
closed lines. For all other cases shown, there are always two
vortices present apart from times where $t =(2 n +1)\pi/2 $, with
$n$ an integer number. (b) Trajectories for the more general initial
condition $\br_1 = (0.9,0.1)$, $\br_2 = (-1.1,0)$.
 \label{Fig:Diploe_Traj}}
\end{figure}

The dynamics changes considerably for a vortex dipole. Assuming the
$q=+1$ vortex is initially located at $(x_0,0)$ and the $q=-1$
vortex at $(-x_0,0)$, the trajectories are given by
$y_{\mathrm{vd}}(t) = \sin(t) (x_0^2 -1)/x_0$ and
$x_{\mathrm{vd}}(t) = \pm \sqrt{x_0^2 - y_{\mathrm{vd}}^2(t)}$.
Examples of these trajectories are shown in
Fig.~\ref{Fig:Diploe_Traj}(a). For $x_0 \geq 1/\sqrt{2}$ these
expressions are always real numbers indicating that the two vortices
do not annihilate each other. For $x_0 = 1$ the vortices remain even
stationary at their initial positions. The situation changes for
$x_0 < 1/\sqrt{2}$. In this case, the vortices collide with each
other at a time $t_a$ given by the first possible solution of
\begin{equation} \label{Eq:Anntimes}
  |\sin(t)| = \left| \frac{x_0^2}{x_0^2-1} \right| \,.
\end{equation}
They annihilate each other and only reappear again at time $t_r$,
where the second possible solution of Eq.~(\ref{Eq:Anntimes}) in the
half-period $ t \in [0,\pi]$ is taken. Between those two times,
there are no vortices present in the condensate.
We also did calculations with more general initial conditions for
the two vortices, located at $\br_j = (x_j,y_j)$, as shown in
Fig.~\ref{Fig:Diploe_Traj}(b). The whole dynamics is still periodic
with a period of $2 \pi$, and after a time of $t = \pi$ the state is
given by $\phi_\mathrm{vd}(x,y,\pi) = -\phi_\mathrm{vd}(-x,-y,0)$,
which means the initial state is, up to an unimportant global phase,
inflected at the origin. For the symmetric stable case with $x_0 =
1$ this might be surprising at the first glance, but is explained by
the fact that the (dimensionless) current density $\mathbf{j}(x,y,t)
= - \ii \left( \psi^\ast \nabla \psi
  - \psi \nabla \psi^\ast \right)/2$ stops at times $t =
  (2n+1)\pi/2$, $n = 1,2,...$,
allowing the two stationary vortices to flip their signs.

The dynamics of a vortex tripole with two vortices of topological
charge $q=+1$ at locations $(x_0,0)$ and $(-x_0,0)$ and one of
charge $q=-1$ at $(0,0)$ is given by
\begin{equation}
\begin{split}\label{Eq:VT_0}
 \phi_{\mathrm{vt}} (x,y,t) =& \eh^{-\frac{x^2+y^2}{2}-4 \ii t} \\
 &\big[x^3+\ii y x^2+\left(y^2-\eh^{2 \ii t} \left(x_0^2-2\right)-2\right) x \\
 &+ \ii y
   \left(y^2+\eh^{2 \ii t} \left(x_0^2+2\right)-2\right)\big] \,.
\end{split}
\end{equation}
For the special case $x_0 = \sqrt{2}$, we find as the zeros of the
wave function $(0,0)$, $(-\sqrt{2},0)$, $(\sqrt{2},0)$, $(\sqrt{2 -
4 \cos(2t)}\sin(2 t), -\sqrt{2 - 4 \cos(2t)} \cos(2 t))$, and
$(-\sqrt{2 - 4 \cos(2t)}\sin(2 t), \sqrt{2 - 4 \cos(2t)} \cos(2
t))$. Although the initial state only contains three vortices,
during the evolution additional vortices are created and annihilated
again, so that at certain times there is a maximum of five vortices
present in the condensate. The whole evolution is periodic with a
period of $\pi$, a property which will be lost for small
interactions. For more details of the vortex dynamics in the ideal
case we refer to the Appendix.


\section{Vortices in a weakly interacting BEC \label{Sec:IA}}

The situation gets more complicated when a finite interaction
$\beta$ of the condensate is taken into account, and exact
analytical solutions are not known. In order to calculate the
dynamics of the vortices we therefore proceed using the Ritz
minimization method \cite{Cohen-Tannoudji-1977}. For small
interactions $\beta \leq 1$ we assume that the Gaussian shape of the
condensate is not changed, but only broadened. For the solutions of
the Gross-Pitaevskii equation we make the ansatz
\begin{equation}
  \psi_{n,m,\beta}(\br,t) = \frac{\eh^{- \ii \mu_{n,m}
  t -\frac{(x^2 +
  y^2)}{2 \sigma^2} }}{\sqrt{2^{n+m} n! m! \pi \sigma^2}}\,  H_n(x/\sigma) H_m(y/\sigma) \,,
\end{equation}
where the functions are normalized to 1 and $\sigma$ is a constant
which takes the broadening into account. This constant is derived by
minimizing the Gross-Pitaevskii energy functional
\cite{Pitaevskii-2003}
\begin{equation}
  E[\psi] = N \int \! \diff \br \, \frac{1}{2} | \nabla \psi|^2 +
\frac{1}{2} \br^2 |\psi|^2 + \frac{\beta}{2} |\psi|^4
\end{equation}
with respect to $\sigma$, where $\psi = \psi_{0,0,\beta}(\br,0)$ and
the energy is given in units of $\hbar \omega$. The minimum is found
for $\sigma^2 = \sqrt{(\beta + 2 \pi)/2 \pi}$. The constants
$\mu_{n,m}$ are derived by putting the ansatz into the GPE
(\ref{Eq:GPE}), multiplying by $\psi^\ast$ and integrating over
space, i.e.,
\begin{equation}
  \mu_{n,m} = \int \! \diff \br \, \frac{1}{2} | \nabla \psi_{n,m,\beta}|^2 +
  \frac{1}{2} \br^2 |\psi_{n,m,\beta}|^2 + \beta |\psi_{n,m,\beta}|^4 \,.
\end{equation}
The wave functions $\psi_{n,m,\beta}(\br,t)$ are then used to expand
the vortex state, which is given by the functions describing the
vortices multiplied by $\psi_{0,0,\beta}$ and a normalization
factor. This factor is, as in the preceding sections, dropped unless
otherwise stated.

We have further assumed that the shape of the vortices does not
change due to the increased interaction. Normally, in a strongly
nonlinear condensate the size of a vortex is given by the coherence
length $\xi$, for which we find (in scaled units) $\xi^2 = \sqrt{\pi
/ 4 \beta}$. However, for our trapped weakly interacting condensate
this length scale is no longer useful, which gets especially
apparent for a vanishing interaction $\beta \to 0$, leading to $\xi
\to \infty$. Instead by using numerical calculations we find that
the vortices are still well described by the functions $p_q$ derived
for $\beta = 0$.
\begin{figure}
\includegraphics[width=8.5cm]{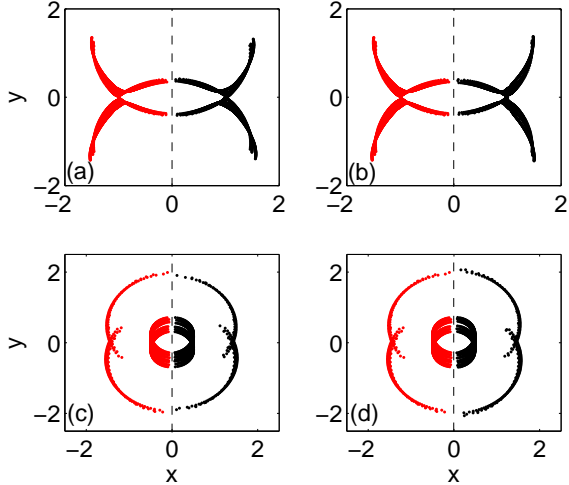}
\caption{(Color online) Comparison of vortex trajectories for
different initial states for a vortex dipole with initial $x_0 = 1$
((a) and (b)) and $x_0 = 0.5$ ((c) and (d)). For $x<0$ the
trajectories for a vortex dipole with initial $p_q$ for $\beta = 1$
is shown, whereas for $x>0$ the trajectories with initial $p_q$ for
$\beta = 0$ ((a) and (c)) and $\beta = 2$ ((b) and (d)) are plotted.
All trajectories are mirror symmetric to the axis where $x=0$
(dashed line). The vortices were evolved for a time interval
$[0,20]$ and a condensate with $\beta = 1$. \label{Fig:InitialCond}}
\end{figure}%
This is illustrated in Figs.~\ref{Fig:InitialCond}(a) and
\ref{Fig:InitialCond}(c), where the trajectories for a vortex dipole
with the analytical $p_q$ for $\beta = 0$ are compared to those with
numerically found $p_q$ for $\beta = 1$. The difference between the
trajectories is negligible. In Figs.~\ref{Fig:InitialCond}(b) and
\ref{Fig:InitialCond}(d) we compare the vortex trajectories for the
factor $p_q$ found for $\beta = 1$ and $\beta = 2$. Again, the
trajectories are almost identical. This shows that a small
perturbation in the shape of the vortices leaves the trajectories
essentially unaffected as long as an initial condition described by
Eq.~(\ref{Eq:InitalData}) is used. We note that
Eq.~(\ref{Eq:InitalData}) only describes initially factorized
vortices, which are not necessarily preferred as \emph{the} initial
states with a given number of vortices. Alternative states can be
considered, and it has been shown that the dynamics for globally
linked, that means nonfactorized, vortices can behave significantly
different from the dynamics of factorized ones
\cite{Crasovan-PRE-2002}. However, a full characterization of such
states is beyond the scope of this paper and we therefore restrict
our considerations to initial states described by
Eq.~(\ref{Eq:InitalData}).

\subsection{Single vortex}
\begin{figure}
\includegraphics[width=8.5cm]{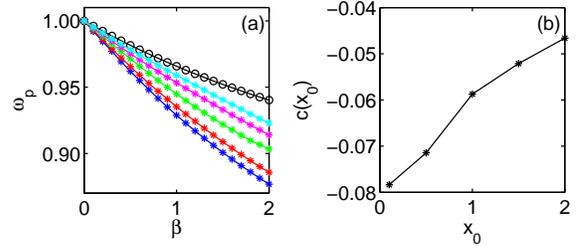}
\caption{(Color online) (a) Precession frequency of a single vortex
versus the interaction strength $\beta$. Stars show $\omega_p$ for a
different initial distance of the vortex from the center, namely
(from bottom to top) $x_0 = 0.1, 0.5, 1.0, 1.5, 2.0$. The circles
show the analytical result given in Eq.~(\ref{Eq:Precession}), lines
are guides to the eye. (b) The coefficients $c(x_0)$ for different
$x_0$, see text. \label{Fig:Precession}}
\end{figure}

For a single vortex initially located at $(x_0,0)$ the expansion
into the solutions $\psi_{n,m,\beta}$ yields
\begin{equation}
\phi_{\mathrm{v}}(x,y,t) = \eh^{-\frac{\ii (7 \beta +16 \pi ) t}{4
\sqrt{2 \pi } \sqrt{\beta +2 \pi }}-\frac{x^2+y^2}{2 \sigma ^2}}
\left(x-\eh^{\frac{\ii (\beta +8 \pi ) t}{4 \sqrt{2 \pi
   } \sqrt{\beta +2 \pi }}} x_0+i y\right)\,,
\end{equation}
which immediately leads to the trajectory
\begin{gather}
  x_{\mathrm{v}}(t) = x_0 \cos \left(\frac{(\beta +8 \pi ) t}{4 \sqrt{2 \pi } \sqrt{\beta +2 \pi
  }}\right) \,,\\
  y_{\mathrm{v}}(t) = x_0 \sin \left(\frac{(\beta +8 \pi ) t}{4 \sqrt{2 \pi } \sqrt{\beta +2 \pi
  }}\right)\,.
\end{gather}
The vortex again moves on an exact circular line, however the time
it needs to complete one circle is increased compared to the
interaction-free case. To be more specific, the precession frequency
is given by
\begin{equation}\label{Eq:Precession}
  \omega_p = \frac{(\beta +8 \pi )}{4 \sqrt{2 \pi } \sqrt{\beta +2 \pi
  }} \,.
\end{equation}
A comparison with numerical results shown in
Fig.~\ref{Fig:Precession}(a) illustrates that the analytical result
describes the trend of a decreasing precession frequency well,
however, the numerical results show a clear dependence on the
distance $x_0$ of the vortex to the center of the trap, which is
missing in the analytical formula. For small interaction $\beta$ the
behavior of the precession frequency can be assumed to be linear
with $\beta$ and follow the curve $\omega_p = 1 + c(x_0)\beta$. For
the analytical formula we get $c_\mathrm{ana}(x_0) = -1/8\pi \approx
-0.04$, whereas the results from the numerics are shown in
Fig.~\ref{Fig:Precession}(b). A trend towards the analytical value
is visible with increasing $x_0$. However, measuring the frequency
for $x_0 > 2$ will become increasingly difficult due to the dilute
condensate density for large distances.

\subsection{Vortex pair}
\begin{figure}
\includegraphics[width=6cm]{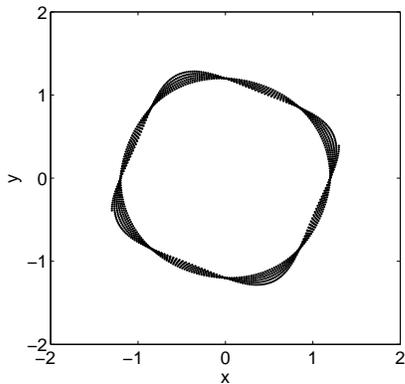}
\caption{Trajectories of a vortex pair deduced from the analytical
formula Eq.~(\ref{Eq:VP_IA}) for $\beta = 2$, $x_0 = 1.2$, and a
time interval of $t \in [0,20]$. The trajectories are no longer
exact circular lines, indicating that both vortices influence each
other. \label{Fig:Pair_IA_Traj}}
\end{figure}

Let us assume that at time $t=0$ the two vortices with topological
charge $q_j = +1$ are located at $(-x_0,0)$ and $(x_0,0)$. The time
evolution of the BEC wave function is given by
\begin{equation} \label{Eq:VP_IA}
\begin{split}
\phi_{\mathrm{vp}} =&   \eh^{-\frac{\ii (137 \beta +384 \pi ) t}{64
\sqrt{2 \pi } \sqrt{\beta +2 \pi }}-\frac{x^2+y^2}{2 \sigma ^2}}
\Big(x^2+2 \ii \eh^{\frac{5 \ii \beta  t}{64 \sqrt{2
   \pi } \sqrt{\beta +2 \pi }}} y x \\
   &-\eh^{\frac{\ii (41 \beta +256 \pi ) t}
   {64 \sqrt{2 \pi } \sqrt{\beta +2 \pi }}} x_0^2-y^2\Big) \,.
\end{split}
\end{equation}
We see that the factor describing the two vortices in general cannot
be factorized as in the interaction-free case, indicating that the
two vortices influence each other. This also becomes evident when
investigating the trajectories of the two vortices. They no longer
move on exact circular lines, but rather on deformed ones as shown
in Fig.~\ref{Fig:Pair_IA_Traj}. The deviation from the exact
circular line gets larger for higher interaction strength $\beta$.

\subsection{Vortex dipole}

\begin{figure}
\includegraphics[width=9cm]{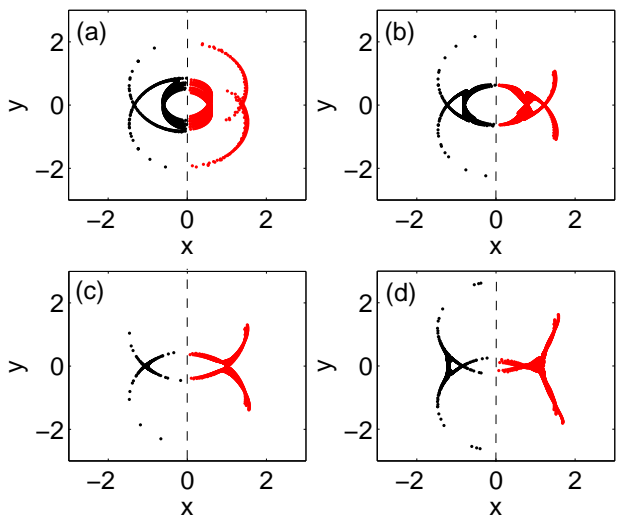}
\caption{(Color online) Trajectories of the two vortices in a vortex
dipole for (a) $x_0 = 0.6$, (b) $x_0 = 0.8$, (c) $x_0 = 1$, and (d)
$x_0 = 1.2$. Analytical results calculated with Eq.~(\ref{Eq:VD_IA})
are only shown for $x<0$, whereas numerical results are only shown
for $x>0$. Both results are mirror symmetric with respect to the
axis where $x=0$ (dashed line). The interaction is chosen as $\beta
= 1$, and the time interval for which the trajectories are shown is
given by $0\leq t \leq 20$. } \label{Fig:traj_b1}
\end{figure}
For symmetric initial conditions, i.e., a negative vortex initially
at position $(-x_0,0)$ and a positive one at position $(x_0,0)$, the
wave function for a vortex dipole is expanded in terms of
$\psi_{n,m,\beta}$ yielding
\begin{equation} \label{Eq:VD_IA}
\begin{split}
\phi_{\mathrm{vd}}(\br,t)=& \eh^{-\frac{3 \ii (115 \beta +256 \pi )
t}{64 \sqrt{2 \pi } \sqrt{\beta +2 \pi }}-\frac{x^2+y^2}{2 \sigma
^2}} \Bigg(2 \ii \eh^{\frac{\ii (233 \beta +512 \pi ) t}{64 \sqrt{2
\pi } \sqrt{\beta +2
   \pi }}} x_0 y  \\
   &+ \eh^{\frac{\ii (249 \beta +640 \pi ) t}{64
 \sqrt{2 \pi } \sqrt{\beta +2 \pi }}} (\sigma -x_0) (\sigma +x_0)
 \\
 &+\eh^{\frac{\ii (13 \beta +24 \pi ) t}{4 \sqrt{2 \pi }
   \sqrt{\beta +2 \pi }}} \left(-\sigma
   ^2+x^2+y^2\right)\Bigg) \,.
\end{split}
\end{equation}

As for the vortex pair the dynamics of the BEC is no longer periodic
due to the interaction, but acquires a more complicated time
dependence. As an example, the trajectories for $\beta = 1$, a time
interval of $t \in [0,20]$, and several initial positions $(\pm
x_0,0)$ are shown in Fig.~\ref{Fig:traj_b1}. The simple trajectories
from the noninteracting case are changed to complicated structures,
which have lost their periodicity.

Comparison between analytical results using Eq.~(\ref{Eq:VD_IA}) and
numerics shows that especially for small times and intermediate
distances $2x_0 \approx 2$ both trajectories agree quite well,
however for certain times the analytical results predict vortices at
positions where there should be no vortices according to numerics.
The differences between analytics and numerics get larger for
increasing interaction $\beta$. Our numerical calculations
furthermore suggest that even for distances $2x_0 > 2 $ there exist
times for which the two vortices annihilate each other and reappear
again, so, for example, at $2x_0 = 3$ and a time around $t = 18$.
This is in contrast to the noninteracting case, where the evolution
was strictly periodic and an annihilation of the vortices not
possible if their initial distance was  $2x_0 > 2/ \sqrt{2}$.

\subsection{Vortex tripole}

For a vortex tripole as introduced in Sec.~\ref{Sec:Ideal} the wave
function describing the time evolution is given by

\begin{widetext}
\begin{equation}\label{Eq:VT_Ia}
\begin{split}
\phi_\mathrm{vt}(x,y,t) =&  \eh^{-\frac{871 \ii \beta t}{128 \sqrt{2
\pi } \sqrt{\beta +2 \pi }}-\frac{\frac{20 \ii \sqrt{2 \pi } t
\sigma ^2}{\sqrt{\beta +2 \pi
   }}+x^2+y^2}{2 \sigma ^2}} \Bigg(\eh^{\frac{3 \ii (369 \beta +1024 \pi ) t}
   {256 \sqrt{2 \pi } \sqrt{\beta +2 \pi }}}
   \left(\sigma ^2 (x+\ii y)-2 \ii
   x (x-\ii y) y\right)              \\
   &+\eh^{\frac{3 \ii (361 \beta +1024 \pi ) t}{256 \sqrt{2 \pi }
   \sqrt{\beta +2 \pi }}} \left(3 \sigma ^2 (x+\ii y)-2 \left(x^3+\ii
   y^3\right)\right)-2 \eh^{\frac{\ii (647 \beta +2048 \pi ) t}{128 \sqrt{2 \pi }
   \sqrt{\beta +2 \pi }}} \left(2 \sigma ^2 (x+\ii y)-(x-\ii y)
   x_0^2\right)\Bigg) \,.
\end{split}
\end{equation}
\end{widetext}
As for the noninteracting case, there always exists a vortex in the
center of the condensate, however we were not able to identify any
initial condition where more than the central vortex are stationary.
This is consistent with the results reported in
Ref.~\cite{Pietilae-PRA-2006}. There it was shown that a stable
vortex tripole, i.e., a configuration of exactly three stationary
vortices, only exists for interactions $\beta \gg 1$, where our
ansatz is no longer valid.

\begin{figure}
\includegraphics[width=8.5cm]{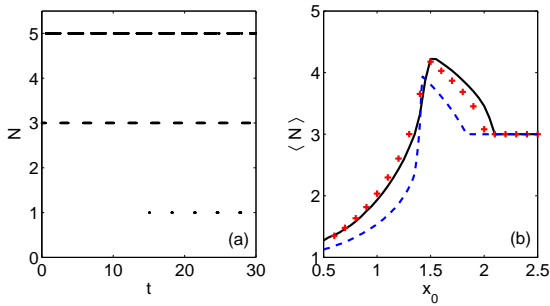}
\caption{(Color online) (a) Number of vortices $ N $ present in the
BEC vs time, with initially $x_0 = 1.5$. (b) Average number of
vortices $\langle N \rangle$ present in the BEC during the time
interval [0,20] for different initial $x_0$. The solid line shows
the results derived from the analytical expression,
Eq.~(\ref{Eq:VT_Ia}), whereas the plus signs show numerical results.
In all cases we have chosen $\beta = 1$, apart from the dashed line,
which shows the results for $\beta = 0$. \label{Fig:NoVortices}}
\end{figure}

The zeros of the wave function Eq.~(\ref{Eq:VT_Ia}) are found
numerically for different initial positions of the vortices. For
short times $t \leq 10$, the time evolution of the trajectories is
quasiperiodic, however this quasiperiodicity is more and more washed
out for longer times. This gets also apparent when the number of
vortices during time is considered. As shown in
Fig.~\ref{Fig:NoVortices}(a), the number of vortices oscillates for
$t \leq 10$ between 3 and 5 for an initial configuration with $x_0 =
1.5$. For later times, however, there are intervals in which the
condensate only exhibits one vortex.

In Fig.~\ref{Fig:NoVortices}(b) we show the average number of
vortices $\langle N \rangle$ in the condensate during the time
interval $[0,20]$ versus the initial positions $x_0$. The results
derived from the numerical evolution of the GPE agree quite well
with the ones from the analytical formula Eq.~(\ref{Eq:VT_Ia}). We
observe that for $x_0 \lesssim 1.5$ the average number of vortices
$\langle N \rangle$ increases with increasing $x_0$, and reaches a
pronounced maximum at $x_0 \approx 1.5$. For larger $x_0$ the
average vortex number decreases again to reach a constant value of 3
for $x_0 \gtrsim 2$. This indicates that for such large distances
the energy within the condensate is too low to spontaneously create
additional vortices. This behavior is qualitatively similar to the
noninteracting case as also indicated in
Fig.~\ref{Fig:NoVortices}(b), where the interacting case tends to
exhibit a higher average vortex number.

\section{Conclusion \label{Sec:Conclusion}}

In the present paper we have investigated the dynamics of vortices
in two-dimensional Bose-Einstein condensates. We have solved the
Gross-Pitaevskii equation analytically for ideal condensates and
have used the Ritz minimization method in order to calculate the
dynamics for small interactions. The latter results were compared to
exact numerical solutions of the condensate dynamics.

For an ideal condensate we have shown that two vortices with the
same topological charge $|q| = 1$ do not influence each other and
behave like two independent, single vortices. This changes as soon
as the interaction within the condensate is taken into account. The
two vortices no longer move on exact circular lines, but rather on a
distorted circular path.

For the case of a vortex dipole, that is two vortices with opposite
topological charge in the condensate, we found that for certain
initial conditions in an ideal condensate the two vortices will
collide, thereby annihilating each other, and reappear again,
whereas they are always separated if the initial distance is large
enough. The trajectories change considerably in an interacting
condensate, and a large initial distance will no longer guarantee
that the two vortices do not annihilate each other at some times.

We also investigated the case where initially there are three
vortices in the BEC, a so-called vortex tripole. In contrast to the
case of only two vortices present at the beginning we found that for
an ideal condensate during the time evolution additional vortices
were created and annihilated, allowing for a maximum of five
vortices in the condensate. Our numerical results showed that this
behavior also persisted in an interacting condensate.

\acknowledgments{This work was supported by the National University
of Singapore Grant No. R-146-000-083-112, the EPSRC (UK) through the
QIP IRC (GR/S82176/01) and the EuroQUAM project EP/E041612/1, and by
the EU through the STREP project OLAQUI. One of the authors (A.K.)
acknowledges financial support from the Keble Association.}


\appendix

\section{Details of the vortex tripole dynamics for the ideal BEC}

\begin{figure}
\includegraphics[width=8cm]{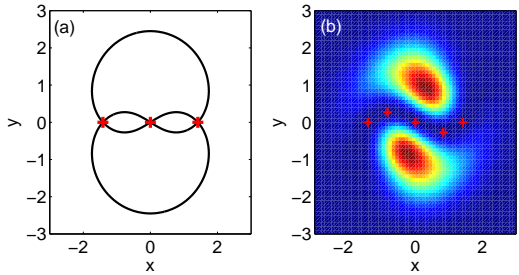}
\caption{(Color online) (a) Trajectories of the two additional
vortices (black lines) for the initial state of a vortex tripole
with $x_0 = \sqrt{2}$, details see text. The three stationary
vortices are indicated by plus signs. (b) Density plot for $t = 0.2
\pi$. Plus signs indicate the position of the five vortices.
\label{Fig:Tripole_Traj}}
\end{figure}%

The expressions given for the vortex trajectories in
Sec.~\ref{Sec:Ideal} do not include information on the topological
charge of the respective vortices. Closer investigations show that
the vortices described by a single trajectory can indeed flip their
signs. To be more precise, we assume that at $t=0$ the central
vortex has a negative charge. At times $t<\pi/6$, there are only
three vortices present in the BEC, which do not move. For $t=\pi/6$,
two new vortices of equal charge $q=-1$ arise in the center of the
BEC flipping the charge of the central vortex in order to keep the
total topological charge $Q$ constant. The two new vortices start to
move out of the center towards the two stationary vortices. When the
two new vortices cross the two stationary ones at $t = \pi/4$, their
charges are flipped as well, such that the moving vortices have now
a positive charge and the stationary ones a negative one. After the
flip, the nonstationary vortices move around the three stationary
vortices as indicated in Fig.~\ref{Fig:Tripole_Traj}(a), cross the
two vortices outside the center again and flipping their charge at
$t = 3 \pi /4$, and finally annihilate at the center at a time
$t=5\pi/6$, such that the original configuration for $t=0$ is
achieved again. This is repeated periodically, with a period of $
\pi$. Investigating the phase of the condensate at the times $t =
\pi/4$ and $3 \pi /4$, where the moving vortices cross the
stationary ones at $(\pm \sqrt{2},0)$, shows that for these times
there is only one vortex present in the condensate, namely the
central one. A density plot of the system for a time $t = 0.2 \pi$
is shown in Fig.~\ref{Fig:Tripole_Traj}(b). The additional vortices
are created in a region of low density, where the energetic cost is
lowest. However, this low density makes the measurement of the
vortices more difficult.

For the case $x_0 \neq \sqrt{2}$ we find a similar behavior of the
vortices. For $x_0 \geq 2$ there will always be three vortices in
the condensate. For smaller values, additional vortices can be
created, and for $x_0 < \sqrt{2}$ we find that for certain time
intervals (i.e., not only for single points in time) there is only
one vortex present in the condensate. Due to the conservation of the
total topological charge this is the minimum number of vortices
present. Calculating the roots of Eq.~(\ref{Eq:VT_0}) shows that the
maximum number of vortices is five.

\bibliography{Vortex}

\begin{thebibliography}{32}
\expandafter\ifx\csname natexlab\endcsname\relax\def\natexlab#1{#1}\fi
\expandafter\ifx\csname bibnamefont\endcsname\relax
  \def\bibnamefont#1{#1}\fi
\expandafter\ifx\csname bibfnamefont\endcsname\relax
  \def\bibfnamefont#1{#1}\fi
\expandafter\ifx\csname citenamefont\endcsname\relax
  \def\citenamefont#1{#1}\fi
\expandafter\ifx\csname url\endcsname\relax
  \def\url#1{\texttt{#1}}\fi
\expandafter\ifx\csname urlprefix\endcsname\relax\def\urlprefix{URL }\fi
\providecommand{\bibinfo}[2]{#2}
\providecommand{\eprint}[2][]{\url{#2}}

\bibitem[{\citenamefont{Matthews et~al.}(1999)\citenamefont{Matthews, Anderson,
  Haljan, Hall, Wieman, and Cornell}}]{Matthews-PRL-1999}
\bibinfo{author}{\bibfnamefont{M.~R.} \bibnamefont{Matthews}},
  \bibinfo{author}{\bibfnamefont{B.~P.} \bibnamefont{Anderson}},
  \bibinfo{author}{\bibfnamefont{P.~C.} \bibnamefont{Haljan}},
  \bibinfo{author}{\bibfnamefont{D.~S.} \bibnamefont{Hall}},
  \bibinfo{author}{\bibfnamefont{C.~E.} \bibnamefont{Wieman}},
  \bibnamefont{and} \bibinfo{author}{\bibfnamefont{E.~A.}
  \bibnamefont{Cornell}}, \bibinfo{journal}{Phys. Rev. Lett.}
  \textbf{\bibinfo{volume}{83}}, \bibinfo{pages}{2498} (\bibinfo{year}{1999}).

\bibitem[{\citenamefont{Madison
  et~al.}(2000{\natexlab{a}})\citenamefont{Madison, Chevy, Wohlleben, and
  Dalibard}}]{Madison-PRL-2000}
\bibinfo{author}{\bibfnamefont{K.~W.} \bibnamefont{Madison}},
  \bibinfo{author}{\bibfnamefont{F.}~\bibnamefont{Chevy}},
  \bibinfo{author}{\bibfnamefont{W.}~\bibnamefont{Wohlleben}},
  \bibnamefont{and} \bibinfo{author}{\bibfnamefont{J.}~\bibnamefont{Dalibard}},
  \bibinfo{journal}{Phys. Rev. Lett.} \textbf{\bibinfo{volume}{84}},
  \bibinfo{pages}{806} (\bibinfo{year}{2000}{\natexlab{a}}).

\bibitem[{\citenamefont{Madison
  et~al.}(2000{\natexlab{b}})\citenamefont{Madison, Chevy, Wohlleben, and
  Dalibard}}]{Madison-JMO-2000}
\bibinfo{author}{\bibfnamefont{K.~W.} \bibnamefont{Madison}},
  \bibinfo{author}{\bibfnamefont{F.}~\bibnamefont{Chevy}},
  \bibinfo{author}{\bibfnamefont{W.}~\bibnamefont{Wohlleben}},
  \bibnamefont{and} \bibinfo{author}{\bibfnamefont{J.}~\bibnamefont{Dalibard}},
  \bibinfo{journal}{J. Mod. Opt.} \textbf{\bibinfo{volume}{47}},
  \bibinfo{pages}{2715} (\bibinfo{year}{2000}{\natexlab{b}}).

\bibitem[{\citenamefont{Abo-Shaeer et~al.}(2001)\citenamefont{Abo-Shaeer,
  Raman, Vogels, and Ketterle}}]{Abo-Shaeer-Science-2001}
\bibinfo{author}{\bibfnamefont{J.~R.} \bibnamefont{Abo-Shaeer}},
  \bibinfo{author}{\bibfnamefont{C.}~\bibnamefont{Raman}},
  \bibinfo{author}{\bibfnamefont{J.~M.} \bibnamefont{Vogels}},
  \bibnamefont{and} \bibinfo{author}{\bibfnamefont{W.}~\bibnamefont{Ketterle}},
  \bibinfo{journal}{Science} \textbf{\bibinfo{volume}{292}},
  \bibinfo{pages}{476} (\bibinfo{year}{2001}).

\bibitem[{\citenamefont{Raman et~al.}(2001)\citenamefont{Raman, Abo-Shaeer,
  Vogels, Xu, and Ketterle}}]{Raman-PRL-2001}
\bibinfo{author}{\bibfnamefont{C.}~\bibnamefont{Raman}},
  \bibinfo{author}{\bibfnamefont{J.~R.} \bibnamefont{Abo-Shaeer}},
  \bibinfo{author}{\bibfnamefont{J.~M.} \bibnamefont{Vogels}},
  \bibinfo{author}{\bibfnamefont{K.}~\bibnamefont{Xu}}, \bibnamefont{and}
  \bibinfo{author}{\bibfnamefont{W.}~\bibnamefont{Ketterle}},
  \bibinfo{journal}{Phys. Rev. Lett.} \textbf{\bibinfo{volume}{87}},
  \bibinfo{pages}{210402} (\bibinfo{year}{2001}).

\bibitem[{\citenamefont{Leanhardt et~al.}(2002)\citenamefont{Leanhardt,
  G{\"{o}}rlitz, Chikkatur, Kielpinski, Shin, Pritchard, and
  Ketterle}}]{Leanhardt-PRL-2002}
\bibinfo{author}{\bibfnamefont{A.~E.} \bibnamefont{Leanhardt}},
  \bibinfo{author}{\bibfnamefont{A.}~\bibnamefont{G{\"{o}}rlitz}},
  \bibinfo{author}{\bibfnamefont{A.~P.} \bibnamefont{Chikkatur}},
  \bibinfo{author}{\bibfnamefont{D.}~\bibnamefont{Kielpinski}},
  \bibinfo{author}{\bibfnamefont{Y.}~\bibnamefont{Shin}},
  \bibinfo{author}{\bibfnamefont{D.~E.} \bibnamefont{Pritchard}},
  \bibnamefont{and} \bibinfo{author}{\bibfnamefont{W.}~\bibnamefont{Ketterle}},
  \bibinfo{journal}{Phys. Rev. Lett} \textbf{\bibinfo{volume}{89}},
  \bibinfo{pages}{190403} (\bibinfo{year}{2002}).

\bibitem[{\citenamefont{Castin et~al.}(2006)\citenamefont{Castin, Hadzibabic,
  Stock, Dalibard, and Stringari}}]{Castin-PRL-2006}
\bibinfo{author}{\bibfnamefont{Y.}~\bibnamefont{Castin}},
  \bibinfo{author}{\bibfnamefont{Z.}~\bibnamefont{Hadzibabic}},
  \bibinfo{author}{\bibfnamefont{S.}~\bibnamefont{Stock}},
  \bibinfo{author}{\bibfnamefont{J.}~\bibnamefont{Dalibard}}, \bibnamefont{and}
  \bibinfo{author}{\bibfnamefont{S.}~\bibnamefont{Stringari}},
  \bibinfo{journal}{Phys. Rev. Lett.} \textbf{\bibinfo{volume}{96}},
  \bibinfo{pages}{040405} (\bibinfo{year}{2006}).

\bibitem[{\citenamefont{Kapale and Dowling}(2005)}]{Kapale-PRL-2005}
\bibinfo{author}{\bibfnamefont{K.~T.} \bibnamefont{Kapale}} \bibnamefont{and}
  \bibinfo{author}{\bibfnamefont{J.~P.} \bibnamefont{Dowling}},
  \bibinfo{journal}{Phys. Rev. Lett.} \textbf{\bibinfo{volume}{95}},
  \bibinfo{pages}{173601} (\bibinfo{year}{2005}).

\bibitem[{\citenamefont{Jackson et~al.}(1999)\citenamefont{Jackson, McCann, and
  Adams}}]{Jackson-PRA-1999}
\bibinfo{author}{\bibfnamefont{B.}~\bibnamefont{Jackson}},
  \bibinfo{author}{\bibfnamefont{J.~F.} \bibnamefont{McCann}},
  \bibnamefont{and} \bibinfo{author}{\bibfnamefont{C.~S.} \bibnamefont{Adams}},
  \bibinfo{journal}{Phys. Rev. A} \textbf{\bibinfo{volume}{61}},
  \bibinfo{pages}{013604} (\bibinfo{year}{1999}).

\bibitem[{\citenamefont{Feder et~al.}(2001)\citenamefont{Feder, Svidzinsky,
  Fetter, and Clark}}]{Feder-PRL-2001}
\bibinfo{author}{\bibfnamefont{D.~L.} \bibnamefont{Feder}},
  \bibinfo{author}{\bibfnamefont{A.~A.} \bibnamefont{Svidzinsky}},
  \bibinfo{author}{\bibfnamefont{A.~L.} \bibnamefont{Fetter}},
  \bibnamefont{and} \bibinfo{author}{\bibfnamefont{C.~W.} \bibnamefont{Clark}},
  \bibinfo{journal}{Phys. Rev. Lett.} \textbf{\bibinfo{volume}{86}},
  \bibinfo{pages}{564} (\bibinfo{year}{2001}).

\bibitem[{\citenamefont{Martikainen et~al.}(2001)\citenamefont{Martikainen,
  Suominen, Santos, Schulte, and Sanpera}}]{Martikainen-PRA-2001}
\bibinfo{author}{\bibfnamefont{J.-P.} \bibnamefont{Martikainen}},
  \bibinfo{author}{\bibfnamefont{K.-A.} \bibnamefont{Suominen}},
  \bibinfo{author}{\bibfnamefont{L.}~\bibnamefont{Santos}},
  \bibinfo{author}{\bibfnamefont{T.}~\bibnamefont{Schulte}}, \bibnamefont{and}
  \bibinfo{author}{\bibfnamefont{A.}~\bibnamefont{Sanpera}},
  \bibinfo{journal}{Phys. Rev. A} \textbf{\bibinfo{volume}{64}},
  \bibinfo{pages}{063602} (\bibinfo{year}{2001}).

\bibitem[{\citenamefont{Andrelczyk et~al.}(2001)\citenamefont{Andrelczyk,
  Brewczyk, Dobrek, Gajda, and Lewenstein}}]{Andrelczyk-PRA-2001}
\bibinfo{author}{\bibfnamefont{G.}~\bibnamefont{Andrelczyk}},
  \bibinfo{author}{\bibfnamefont{M.}~\bibnamefont{Brewczyk}},
  \bibinfo{author}{\bibfnamefont{{\L}.}~\bibnamefont{Dobrek}},
  \bibinfo{author}{\bibfnamefont{M.}~\bibnamefont{Gajda}}, \bibnamefont{and}
  \bibinfo{author}{\bibfnamefont{M.}~\bibnamefont{Lewenstein}},
  \bibinfo{journal}{Phys. Rev. A} \textbf{\bibinfo{volume}{64}},
  \bibinfo{pages}{043601} (\bibinfo{year}{2001}).

\bibitem[{\citenamefont{P{\'{e}}rez-Garc{\'{\i}}a
  et~al.}(2007)\citenamefont{P{\'{e}}rez-Garc{\'{\i}}a, Garc{\'{\i}}a-March,
  and Ferrando}}]{Perez-Garcia-PRA-2007}
\bibinfo{author}{\bibfnamefont{V.~M.} \bibnamefont{P{\'{e}}rez-Garc{\'{\i}}a}},
  \bibinfo{author}{\bibfnamefont{M.~A.} \bibnamefont{Garc{\'{\i}}a-March}},
  \bibnamefont{and} \bibinfo{author}{\bibfnamefont{A.}~\bibnamefont{Ferrando}},
  \bibinfo{journal}{Phys. Rev. A} \textbf{\bibinfo{volume}{75}},
  \bibinfo{pages}{033618} (\bibinfo{year}{2007}).

\bibitem[{\citenamefont{Crasovan et~al.}(2003)\citenamefont{Crasovan,
  Vekslerchik, P{\'{e}}rez-Garc{\'{\i}}a, Torres, Mihalache, and
  Torner}}]{Crasovan-PRA-2003}
\bibinfo{author}{\bibfnamefont{L.-C.} \bibnamefont{Crasovan}},
  \bibinfo{author}{\bibfnamefont{V.}~\bibnamefont{Vekslerchik}},
  \bibinfo{author}{\bibfnamefont{V.~M.}
  \bibnamefont{P{\'{e}}rez-Garc{\'{\i}}a}},
  \bibinfo{author}{\bibfnamefont{J.~P.} \bibnamefont{Torres}},
  \bibinfo{author}{\bibfnamefont{D.}~\bibnamefont{Mihalache}},
  \bibnamefont{and} \bibinfo{author}{\bibfnamefont{L.}~\bibnamefont{Torner}},
  \bibinfo{journal}{Phys. Rev. A} \textbf{\bibinfo{volume}{68}},
  \bibinfo{pages}{063609} (\bibinfo{year}{2003}).

\bibitem[{\citenamefont{M{\"o}tt{\"o}nen
  et~al.}(2005)\citenamefont{M{\"o}tt{\"o}nen, Virtanen, Isoshima, and
  Salomaa}}]{Moettoenen-PRA-2005}
\bibinfo{author}{\bibfnamefont{M.}~\bibnamefont{M{\"o}tt{\"o}nen}},
  \bibinfo{author}{\bibfnamefont{S.~M.~M.} \bibnamefont{Virtanen}},
  \bibinfo{author}{\bibfnamefont{T.}~\bibnamefont{Isoshima}}, \bibnamefont{and}
  \bibinfo{author}{\bibfnamefont{M.~M.} \bibnamefont{Salomaa}},
  \bibinfo{journal}{Phys. Rev. A} \textbf{\bibinfo{volume}{71}},
  \bibinfo{pages}{033626} (\bibinfo{year}{2005}).

\bibitem[{\citenamefont{Pietil{\"a} et~al.}(2006)\citenamefont{Pietil{\"a},
  M{\"o}tt{\"o}nen, Isoshima, Huhtam{\"a}ki, and Virtanen}}]{Pietilae-PRA-2006}
\bibinfo{author}{\bibfnamefont{V.}~\bibnamefont{Pietil{\"a}}},
  \bibinfo{author}{\bibfnamefont{M.}~\bibnamefont{M{\"o}tt{\"o}nen}},
  \bibinfo{author}{\bibfnamefont{T.}~\bibnamefont{Isoshima}},
  \bibinfo{author}{\bibfnamefont{J.~A.~M.} \bibnamefont{Huhtam{\"a}ki}},
  \bibnamefont{and} \bibinfo{author}{\bibfnamefont{S.~M.~M.}
  \bibnamefont{Virtanen}}, \bibinfo{journal}{Phys. Rev. A}
  \textbf{\bibinfo{volume}{74}}, \bibinfo{pages}{023603}
  (\bibinfo{year}{2006}).

\bibitem[{\citenamefont{Zhou and Zhai}(2004)}]{Zhou-PRA-2004}
\bibinfo{author}{\bibfnamefont{Q.}~\bibnamefont{Zhou}} \bibnamefont{and}
  \bibinfo{author}{\bibfnamefont{H.}~\bibnamefont{Zhai}},
  \bibinfo{journal}{Phys. Rev. A} \textbf{\bibinfo{volume}{70}},
  \bibinfo{pages}{043619} (\bibinfo{year}{2004}).

\bibitem[{\citenamefont{Rubinstein and Pismen}(1994)}]{Rubinstein-PysicaD-1994}
\bibinfo{author}{\bibfnamefont{B.~Y.} \bibnamefont{Rubinstein}}
  \bibnamefont{and} \bibinfo{author}{\bibfnamefont{L.~M.}
  \bibnamefont{Pismen}}, \bibinfo{journal}{Physica D}
  \textbf{\bibinfo{volume}{78}}, \bibinfo{pages}{1} (\bibinfo{year}{1994}).

\bibitem[{\citenamefont{Pismen}(1999)}]{Pismen-1999}
\bibinfo{author}{\bibfnamefont{L.~M.} \bibnamefont{Pismen}},
  \emph{\bibinfo{title}{Vortices in Nonlinear Fields: From Liquid Crystals to
  Superfluids, From Non-Equilibrium Patterns to Cosmic Strings}}
  (\bibinfo{publisher}{Oxford University Press}, \bibinfo{address}{New York},
  \bibinfo{year}{1999}), \bibinfo{edition}{1st} ed.

\bibitem[{\citenamefont{Anglin}(2002)}]{Anglin-PRA-2002}
\bibinfo{author}{\bibfnamefont{J.~R.} \bibnamefont{Anglin}},
  \bibinfo{journal}{Phys. Rev. A} \textbf{\bibinfo{volume}{65}},
  \bibinfo{pages}{063611} (\bibinfo{year}{2002}).

\bibitem[{\citenamefont{Molina-Terriza
  et~al.}(2001)\citenamefont{Molina-Terriza, Torner, Wright,
  Garc{\'{\i}}a-Ripoll, and P{\'{e}}rez-Garc{\'{\i}}a}}]{Terriza-OptLett-2001}
\bibinfo{author}{\bibfnamefont{G.}~\bibnamefont{Molina-Terriza}},
  \bibinfo{author}{\bibfnamefont{L.}~\bibnamefont{Torner}},
  \bibinfo{author}{\bibfnamefont{E.~M.} \bibnamefont{Wright}},
  \bibinfo{author}{\bibfnamefont{J.~J.} \bibnamefont{Garc{\'{\i}}a-Ripoll}},
  \bibnamefont{and} \bibinfo{author}{\bibfnamefont{V.~M.}
  \bibnamefont{P{\'{e}}rez-Garc{\'{\i}}a}}, \bibinfo{journal}{Opt. Lett.}
  \textbf{\bibinfo{volume}{26}}, \bibinfo{pages}{1601} (\bibinfo{year}{2001}).

\bibitem[{\citenamefont{Roux}(2004)}]{Roux-OC-2004}
\bibinfo{author}{\bibfnamefont{F.~S.} \bibnamefont{Roux}},
  \bibinfo{journal}{Opt. Commun.} \textbf{\bibinfo{volume}{234}},
  \bibinfo{pages}{63} (\bibinfo{year}{2004}).

\bibitem[{\citenamefont{Bao et~al.}(2003)\citenamefont{Bao, Jaksch, and
  Markovich}}]{Bao-JCP-2003}
\bibinfo{author}{\bibfnamefont{W.}~\bibnamefont{Bao}},
  \bibinfo{author}{\bibfnamefont{D.}~\bibnamefont{Jaksch}}, \bibnamefont{and}
  \bibinfo{author}{\bibfnamefont{P.~A.} \bibnamefont{Markovich}},
  \bibinfo{journal}{J. Comput. Phys.} \textbf{\bibinfo{volume}{187}},
  \bibinfo{pages}{318} (\bibinfo{year}{2003}).

\bibitem[{\citenamefont{Bao and Zhang}(2005)}]{Bao-MMMAS-2005}
\bibinfo{author}{\bibfnamefont{W.}~\bibnamefont{Bao}} \bibnamefont{and}
  \bibinfo{author}{\bibfnamefont{Y.}~\bibnamefont{Zhang}},
  \bibinfo{journal}{Math. Models Meth. Appl. Sci.}
  \textbf{\bibinfo{volume}{15}}, \bibinfo{pages}{1863} (\bibinfo{year}{2005}).

\bibitem[{\citenamefont{Bao and Jaksch}(2003)}]{Bao-SIAM-2003}
\bibinfo{author}{\bibfnamefont{W.}~\bibnamefont{Bao}} \bibnamefont{and}
  \bibinfo{author}{\bibfnamefont{D.}~\bibnamefont{Jaksch}},
  \bibinfo{journal}{SIAM J. Numer. Anal.} \textbf{\bibinfo{volume}{41}},
  \bibinfo{pages}{1406} (\bibinfo{year}{2003}).

\bibitem[{\citenamefont{Berezinskii}(1972)}]{Berezinskii-JETP-1972}
\bibinfo{author}{\bibfnamefont{V.~L.} \bibnamefont{Berezinskii}},
  \bibinfo{journal}{Sov. Phys. JETP} \textbf{\bibinfo{volume}{34}},
  \bibinfo{pages}{610} (\bibinfo{year}{1972}).

\bibitem[{\citenamefont{Kosterlitz and Thouless}(1973)}]{Kosterlitz-JPC-1973}
\bibinfo{author}{\bibfnamefont{J.~M.} \bibnamefont{Kosterlitz}}
  \bibnamefont{and} \bibinfo{author}{\bibfnamefont{D.~J.}
  \bibnamefont{Thouless}}, \bibinfo{journal}{J. Phys. C}
  \textbf{\bibinfo{volume}{6}}, \bibinfo{pages}{1181} (\bibinfo{year}{1973}).

\bibitem[{\citenamefont{Hadzibabic et~al.}(2006)\citenamefont{Hadzibabic,
  Kr{\"u}ger, Cheneau, Battelier, and Dalibard}}]{Hadzibabic-Nature-2006}
\bibinfo{author}{\bibfnamefont{Z.}~\bibnamefont{Hadzibabic}},
  \bibinfo{author}{\bibfnamefont{P.}~\bibnamefont{Kr{\"u}ger}},
  \bibinfo{author}{\bibfnamefont{M.}~\bibnamefont{Cheneau}},
  \bibinfo{author}{\bibfnamefont{B.}~\bibnamefont{Battelier}},
  \bibnamefont{and} \bibinfo{author}{\bibfnamefont{J.}~\bibnamefont{Dalibard}},
  \bibinfo{journal}{Nature (London)} \textbf{\bibinfo{volume}{441}},
  \bibinfo{pages}{1118} (\bibinfo{year}{2006}).

\bibitem[{\citenamefont{Kr{\"u}ger et~al.}(2007)\citenamefont{Kr{\"u}ger,
  Hadzibabic, and Dalibard}}]{Krueger-2007}
\bibinfo{author}{\bibfnamefont{P.}~\bibnamefont{Kr{\"u}ger}},
  \bibinfo{author}{\bibfnamefont{Z.}~\bibnamefont{Hadzibabic}},
  \bibnamefont{and} \bibinfo{author}{\bibfnamefont{J.}~\bibnamefont{Dalibard}},
  \bibinfo{journal}{Phys. Rev. Lett.} \textbf{\bibinfo{volume}{99}},
  \bibinfo{pages}{040402} (\bibinfo{year}{2007}).

\bibitem[{\citenamefont{Pitaevskii and Stringari}(2003)}]{Pitaevskii-2003}
\bibinfo{author}{\bibfnamefont{L.}~\bibnamefont{Pitaevskii}} \bibnamefont{and}
  \bibinfo{author}{\bibfnamefont{S.}~\bibnamefont{Stringari}},
  \emph{\bibinfo{title}{Bose-Einstein Condensation}}
  (\bibinfo{publisher}{Clarendon}, \bibinfo{address}{Oxford},
  \bibinfo{year}{2003}).

\bibitem[{\citenamefont{Cohen-Tannoudji
  et~al.}(1977)\citenamefont{Cohen-Tannoudji, Diu, and
  Lalo{\"{e}}}}]{Cohen-Tannoudji-1977}
\bibinfo{author}{\bibfnamefont{C.}~\bibnamefont{Cohen-Tannoudji}},
  \bibinfo{author}{\bibfnamefont{B.}~\bibnamefont{Diu}}, \bibnamefont{and}
  \bibinfo{author}{\bibfnamefont{F.}~\bibnamefont{Lalo{\"{e}}}},
  \emph{\bibinfo{title}{Quantum Mechanics}}
  (\bibinfo{publisher}{Wiley-{I}nterscience}, \bibinfo{address}{New York},
  \bibinfo{year}{1977}).

\bibitem[{\citenamefont{Crasovan et~al.}(2002)\citenamefont{Crasovan,
  Molina-Terriza, Torres, Torner, P{\'{e}}rez-Garc{\'{\i}}a, and
  Mihalache}}]{Crasovan-PRE-2002}
\bibinfo{author}{\bibfnamefont{L.-C.} \bibnamefont{Crasovan}},
  \bibinfo{author}{\bibfnamefont{G.}~\bibnamefont{Molina-Terriza}},
  \bibinfo{author}{\bibfnamefont{J.~P.} \bibnamefont{Torres}},
  \bibinfo{author}{\bibfnamefont{L.}~\bibnamefont{Torner}},
  \bibinfo{author}{\bibfnamefont{V.~M.}
  \bibnamefont{P{\'{e}}rez-Garc{\'{\i}}a}}, \bibnamefont{and}
  \bibinfo{author}{\bibfnamefont{D.}~\bibnamefont{Mihalache}},
  \bibinfo{journal}{Phys. Rev. E} \textbf{\bibinfo{volume}{66}},
  \bibinfo{pages}{036612} (\bibinfo{year}{2002}).

\end{thebibliography}

\end{document}